\documentclass[aps,prd,reprint,nofootinbib,showkeys,showpacs,superscriptaddress,longbibliography,floatfix]{revtex4-1}

\pdfoutput=1

\usepackage{amsmath}
\usepackage{amssymb}
\usepackage[american]{babel}
\usepackage{booktabs}
\usepackage{dcolumn}  
\usepackage[T1]{fontenc}  
\usepackage{graphicx}  
\usepackage[]{hyperref}  
\usepackage[utf8]{inputenc}  
\usepackage{url}
\usepackage[dvipsnames]{xcolor}
\usepackage{float}

\usepackage{tikz}

\usetikzlibrary{arrows,shapes,positioning,shadows,trees}

\tikzset{
	basic/.style  = {draw, text width=2cm, drop shadow, font=\sffamily, rectangle},
	root/.style   = {basic, rounded corners=2pt, thin, align=center,
		fill=green!30},
	level 2/.style = {basic, rounded corners=6pt, thin,align=center, fill=green!60,
		text width=8em},
	level 3/.style = {basic, thin, align=left, fill=pink!60, text width=6.5em}
}

\usetikzlibrary{arrows,decorations.pathmorphing,backgrounds,fit,positioning,shapes.symbols,chains}

\newcolumntype{d}[1]{D{.}{.}{#1}}
\newcolumntype{v}[1]{D{,}{,\ }{#1}}

\makeatletter

\newcommand{\Rmnum}[1]{\expandafter\@slowromancap\romannumeral #1@}
\makeatother

\renewcommand {\arraystretch}{1.3}

\hypersetup{
	citecolor = red,
	linkcolor = blue,
	urlcolor = magenta
}

\begin{document}

\title{Double-well Inflation: observational constraints and theoretical implications}

\author{G. Rodrigues}
\email{gabrielrodrigues@on.br}
\affiliation{Observatório Nacional, 20921-400, Rio de Janeiro, RJ, Brasil}
\affiliation{Departamento de F{\'\i}sica, Universidad de C\'ordoba,
Campus Universitario de Rabanales, Ctra. N-IV Km.~396, E-14071 C\'ordoba, Spain}

\author{J. G. Rodrigues}
\email{jamersoncg@gmail.com}
\affiliation{Observatório Nacional, 20921-400, Rio de Janeiro, RJ, Brasil}

\author{F. B. M. dos Santos}
\email{fbmsantos@on.br}
\affiliation{Observatório Nacional, 20921-400, Rio de Janeiro, RJ, Brasil}

\author{J. S. Alcaniz}
\email{alcaniz@on.br}
\affiliation{Observatório Nacional, 20921-400, Rio de Janeiro, RJ, Brasil}


\begin{abstract}
We revisit the small field double-well inflationary model and investigate its observational viability in light of the current Cosmic Microwave Background data. In particular, considering scenarios with minimal and nonminimal coupling between the inflaton field and the Ricci scalar, we perform a Monte Carlo Markov chain analysis to probe the model's parameter space. We also investigate the consequences of the cosmological results in the canonical type-I seesaw mechanism context and obtain constraints on the vacuum expectation value of the inflaton field, together with the amplitude of the coupling to gravity in the nonminimal case. We employ a Bayesian procedure to compare the model's predictions with the Starobinsky inflationary scenario and find a strong statistical preference for the latter against the minimal and nonminimal coupled double-well scenario.
\end{abstract}

\vskip 1.0cm



\maketitle

\vskip 1.0cm


\section{Introduction}

The inflationary scenario provides a mechanism to explain the initial conditions of the observable universe~\cite{Starobinsky:1980te,Guth:1980zm,Linde:1981mu}. In the standard picture, the cosmic acceleration is driven by the dynamics of a scalar field, the inflaton, through the slow-roll mechanism. Quantum fluctuations are generated in the chaotic regime, inducing curvature perturbations and, ultimately, the fluctuations present in the Cosmic Microwave Background (CMB)~\cite{Ade:2018gkx,Aghanim:2018eyx}, as well as the Large-Scale Structures (LSS) formation~\cite{Ross:2014qpa,Alam:2016hwk,Scolnic:2017caz}.

In the previous decades, there has been an extensive effort to determine which inflationary scenario best describes the cosmological data~\cite{Martin:2013tda}. One particularly simple realization arises from the Double-Well potential\cite{Linde:1994wt,Vilenkin:1994pv,Garcia-Bellido:1996yji,Green:1996xe,Martin:2013tda}, first introduced as a toy symmetry-breaking model~\cite{Goldstone:1961eq}, with the vacuum expectation value (vev) $v_\phi$ setting the symmetry-breaking scale. For a sufficiently large $v_\phi$ value, one expects a flat region for amplitudes of the field close enough to the origin of the potential, allowing for slow-roll inflation. Despite the slow-roll analysis pointing to tension with current data \cite{Planck:2018jri,Planck:2018vyg}, extensions to the canonical scenario may displace the model predictions to the sweet spot of the constrained parameters. This becomes particularly true when a non-minimal coupling between the inflaton field and gravity is evoked, as demonstrated in \cite{Linde:2011nh,Bostan:2018evz}. Nevertheless, a complete, updated statistical analysis of these models is still lacking in the literature.

From another perspective, it is instructive to investigate which other possible roles the inflaton may assume
during the subsequent evolution of the Universe. One interesting conjecture relates the inflaton field to the seesaw mechanism. The seesaw mechanism, proposed in the late '70s, hatches as an elegant description of neutrino masses' origin~\cite{Minkowski:1977sc,Yanagida:1979as,GellMann:1980vs,Mohapatra:1979ia,Schechter:1980gr}. The high-energy scale $M$ defines the lepton number breaking ($\Delta L = 2$), interweaving its diverse realizations. The canonical type-I seesaw includes three sterile fermions,  $N_i$, and one singlet scalar, $\phi$, to the standard model of fundamental particles (SM) content to generate $M$ dynamically. The high energies associated with the lepton number breaking scale $M$ build a challenging scenario for probing $\phi$, taking into account its production at collider experiments or even its off-shell effects on the flavor physics. Still, such phenomena can leave a singular signature in the primordial Universe when related to the inflationary dynamics.

In this work, we investigate the observational viability of the Double-Well inflation in the face of the current CMB data. In particular, we performed a Monte Carlo Markov Chain analysis to apply the Bayesian parameter selection procedure to constrain the usual cosmological parameters and the inflaton vev $v_\phi$. We also explore the impact of a non-zero, non-minimal coupling of the inflaton to gravity, with the strength given by a parameter $\xi$, to establish possible correlations between values
of the coupling and the remaining cosmological parameters. Furthermore, we discuss the physical consequences of associating the inflaton field to the scale of the lepton number breaking in the type-I seesaw mechanisms, $M=v_\phi$. In particular, we consider the effects of the sterile sector on possible rare lepton decays. 

This paper is organized as follows: Sections II and II review the Double-Well model and present our methodology and results, respectively. Section IV details
the type-I Seesaw mechanism and how it relates to our results, while Section V is dedicated to the discussion and our concluding remarks.


\section{The Model} \label{sec:2}

In the Double-Well scenario, the slow-roll predictions are obtained from the prototype symmetry breaking potential,
\begin{equation}
    V(\phi) = \frac{\lambda}{4}(\phi^2 - v^2_\phi)^2\;,
    \label{eq:2.1}
\end{equation}
where $\lambda$ is a dimensionless coupling constant, and  the vev $v_\phi$ that determines the minimum of the potential. In what follows, we explore the possibility of a flat region for small values of the inflaton field, $\phi < v_\phi$, characteristic of small field inflation scenarios \cite{Kolb:1999ar}. Such configuration can also be understood as a particular case of the Hilltop Inflation ($p=2$) (see e.g. \cite{Boubekeur:2005zm,Tzirakis:2007bf}).

The study of the inflationary dynamics follows from the slow-roll parameters,
\begin{equation}
    \epsilon = \frac{M^2_{P}}{2}\left(\frac{ V^{\prime}}{ V}\right)^2 \quad \text{and} \quad \eta  = M^2_{P}\left( \frac{V^{\prime \prime}}{V } \right),
\end{equation}
where $^\prime$ indicates the derivative with respect to the inflaton field $\phi$ and $M_{P}=\frac{1}{\sqrt{8\pi G}}$ stands for the reduced Planck mass. The accelerated expansion occurs mainly in the slow-roll regime where $\epsilon, \, \eta \ll 1$, and continues until $\epsilon, \, \eta \sim 1$. The standard approach quantifies the cosmological perturbations according to the power-law expansion of the primordial power spectrum. For scalar perturbations,
\begin{equation}
    \ln{\frac{P_\mathcal{R}(k)}{P_\mathcal{R}(k_\star)}} = (n_s-1) \ln\left({\frac{k}{k_\star}}\right) + \frac{\alpha_s}{2}\ln\left({\frac{k}{k_\star}}\right)^2 + \ldots , \label{Eq:06}
\end{equation}
and similarly for the tensor perturbations, where the ellipsis represents high order in the logarithm and $k_\ast$ is the pivot scale. Deep inside the slow-roll regime, the inflationary observables and the slow-roll parameters are associated in a simple way. In particular, the predictions for the scalar tilt $n_s$ and the tensor-to-scalar ratio $r$ are computed via:
\begin{equation}
    n_s=1-6\epsilon_\star+2\eta_\star \quad {\rm{and}} \quad r=16\epsilon_\star,
\end{equation}
where the subscript $_\star$ denotes that predictions are evaluated for the amplitude $\phi = \phi_\star$, associated with the time at which the pivot scale crosses the Hubble horizon. The Planck Collaboration \cite{Planck:2018jri} estimates $n_s=0.9651\pm0.0041$ at $68\%$ confidence level (C.L.). An upper limit is also obtained for the tensor-to-scalar ratio, $r < 0.06$, at 95\% (C.L.).

To complete the description of the scalar perturbations one needs to specify the amplitude of scalar perturbations,
\begin{equation}
A_s =P_\mathcal{R}(k_\star) =\left.\frac{V}{24M^4_P\pi^2\epsilon}\right|_{\phi=\phi_\star}.
\label{eq:2.5}
\end{equation}
The Planck Collaboration has set $\ln\left(10^{10}A_s\right)=3.044\pm 0.014$, also at $68\%$ (C.L.). Its explicit relation to $V$ allows to obtain a correlation between the free parameters of the potential. Particularly for the Double-Well, one may have $\lambda(v_\phi)$ or $\lambda(v_\phi,\xi)$, where the latter stands for the non-minimal scenario.

\begin{figure}[!t]
\centering
\includegraphics[width=\columnwidth]{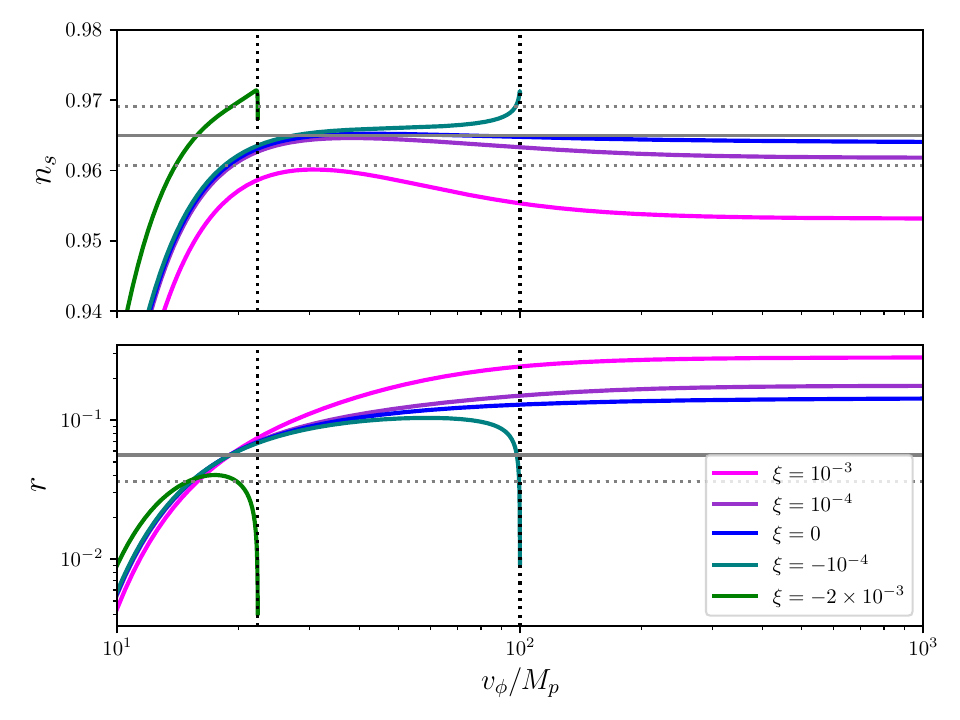}
\caption{$n_s$ and $r$ as a function of the vev $v_\phi/M_P$ for different values of the non-minimal coupling parameter $\xi$, including the minimally coupled model (blue line). In the upper panel, the horizontal lines refer to the Planck constraint of $n_s=0.9649\pm 0.0042$ (at 68\% C.L.), while in the lower panel, the horizontal lines refer to the upper limit on the tensor-to-scalar ratio from BK15 and BK18 of $r<0.056$ and $r<0.036$ (at 95\% C.L.), respectively. The dotted vertical lines indicate the maximum value of $v_\phi/M_P$ for which the slow-roll regime is valid, when the coupling is negative. We fixed the number of e-folds before the end of inflation to $N_\star=55$.}
\label{fig:1}
\end{figure}

For the minimally coupled case, the inflationary predictions of the potential (\ref{eq:2.1}) are known, and Fig.~\ref{fig:1} depicts its behaviour in the the solid blue lines. Note that While a transplanckian vev ($v_\phi/M_P>15$) is necessary for $n_s$ to lie into the viable region allowed by Planck, relatively high values for the tensor-to-scalar ratio $r$ are obtained as consequence, assigning a tension between the predictions of the model and the recent data from the B mode polarization of the CMB as obtained by BICEP/Keck Array telescopes \cite{BICEP2:2018kqh,BICEP:2021xfz}. This can motivate a variety of extensions to the standard inflationary picture, including interacting neutrinos \cite{roy22,cuesta23, cuesta22,barryman23,escudero20,Bostan_2024}, inflation with thermal effects \cite{Berera:1995ie,Berera:2008ar,Bartrum:2013fia}, and non-minimal couplings of the scalar field with the gravitational sector \cite{Bostan:2018evz,PhysRevD.31.3046}. This last avenue has been explored for decades, and it has been shown that for a variety of models that are excluded when considered in the minimal setting, a non-minimal coupling can be able to restore concordance with data \cite{Starobinsky:1980te,Spokoiny:1984bd,Salopek:1988qh,Bezrukov:2007ep}.

Here, we test this approach by considering a gravitational sector enhanced by a non-minimal coupling with the inflaton of the form,
\begin{equation}
    S_J \supset \int d^4x \sqrt{-g}\frac{M^2_P}{2} \Omega^2(\phi) R,
    \label{eq:2.6}
\end{equation}
where $\Omega^2(\phi) = 1 + \xi \phi^2/M^2_P$ and $R$ is the Ricci scalar. When the non-minimal coupling constant $\xi$ equals zero, we recover the minimally coupled model. By using the formalism applied in previous works \cite{Campista:2017ovq,Rodrigues:2020dod,Rodrigues:2021txa,Rodrigues:2023kiz,Santos:2023bnu}, we can readily compute the predictions for the spectral index $n_s$ and the tensor-to-scalar ratio $r$ as shown in fig. \ref{fig:1}. We plot both $n_s$ and $r$ as a function of the vev $v_\phi/M_P$, and we immediately see that when $\xi<0$, there is an upper limit on the vev. This result emerges from the condition that the effective gravitational coupling, settled by $\Omega^2(\phi)$, remains positive during the whole field region where inflation takes place. In the Einstein frame, this limit translates into a infinitely large potential $V_E(\phi)\equiv V(\phi)/\Omega^4$ when the inflaton $\phi$ approaches $M_P/\sqrt{|\xi|}$, which is not suitable for the slow-roll inflation \cite{Linde:2011nh,Reyimuaji:2020goi}.

We notice how the sign of $\xi$ makes all the difference in the ability of the extended model to provide predictions that agree with the current constraints. For $\xi>0$, the predictions worsen as $n_s$ becomes too low while $r$ increases (as shown by the purple curve). Conversely, a negative $\xi$ can guarantee the agreement, as the curves for $n_s$ can pass the viable region while $r$ can be lower than the current upper limits on $r$. This, in turn, also results in a significant decrease in the maximum viable value for $v_\phi/M_P$, as seen when we compare the teal curve ($\xi=-10^{-4}$) and the green one ($\xi=-2\times 10^{-3}$). As already pointed out in \cite{Linde:2011nh}, in the limit of $|\xi|v^2 \rightarrow M^2_P$, the predictions of the small field Double-Well scenario coincide with the canonical Higgs Inflation with $\xi \gg 1$.

\begin{figure}[!t]
\centering
\includegraphics[width=\columnwidth]{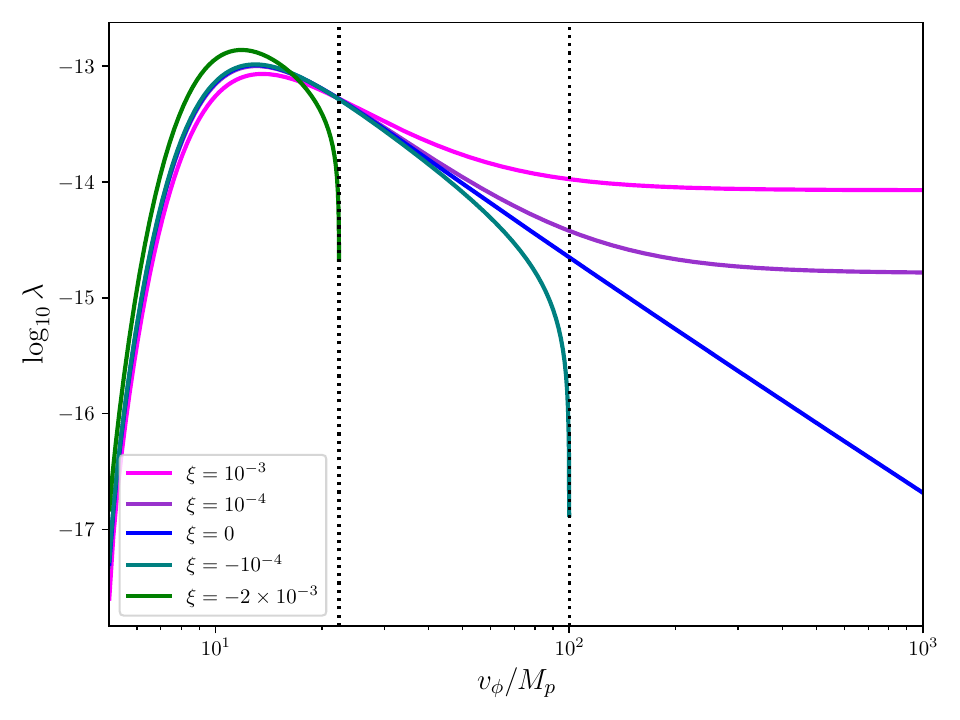}
\caption{$v_\phi$ vs $\log_{10}\lambda$ for $N_\star=55$.}

\label{fig:2}
\end{figure}

From the amplitude of scalar perturbations \eqref{eq:2.5}, one can constrain one of the free parameters of the inflationary potential once the Planck value for $A_s$ is considered. In Fig.~\eqref{fig:2} we present the results in the $v_\phi$ vs $\log_{10}(\lambda)$ plane. Again, the dotted vertical lines denote the upper limit on $v_\phi$ allowed by a negative $\xi$.

\section{Methodology and Results} \label{sec:3}

We perform a comprehensive numerical analysis to constrain the cosmological parameters of the models above discussed. As we also seek to determine the statistical preference for a given model through a robust criterion, we use the \texttt{cobaya} sampler \cite{Torrado:2020dgo,2019ascl.soft10019T}, whose Monte Carlo Markov Chains (MCMC) produced will be analysed by the MCEvidence \texttt{Python} code \cite{Heavens:2017afc} to compute the Bayesian evidence for each model. The theoretical input will be given through an implementation of the double-well potential on the Code for Anisotropies of the Microwave Background (CAMB) \cite{Lewis:1999bs,2012JCAP...04..027H}. For this analysis, we employ a model with the following free parameters: the baryon density parameter $\omega_b$, the cold dark matter density parameter $\omega_{cdm}$, the reionization optical depth $\tau$, and the angular scale of the sound horizon at the drag epoch $\theta_s$. Within the Double-Well model, $n_s$ and $r$ are both derived parameters, as well as $\lambda$, determined from eq. (\ref{eq:2.5}); as a consequence, both $v_\phi/M_P$ and $\xi$ are free to vary. In this manner, we choose to vary the parameter $\log_{10}v_\phi/M_P$ and leave a free $\xi$ or set $\xi=0$ in the non-minimally and minimally coupled models, respectively.

As for the priors of the new parameters, we note that the model supports a weaker coupling with gravity, of order $|\xi|\sim 10^{-3}-10^{-2}$. We then set $\xi:[-0.05,0.05]$, as it is consistent with our initial tests. For $\log_{10}v_\phi/M_P$, we leave the lower limit above the one allowed for the model to achieve slow-roll inflation, being $\log_{10}v_\phi/M_P=0.5$, while we set $\log_{10}v_\phi/M_P=2$ as a general upper limit, but being controlled by the condition on $\xi$ described above. It's worth mentioning that we fix the sum of neutrino masses at 0.06 eV and utilize the pivot scale set by Planck for the primordial spectrum, $k_{\star}=0.05$ Mpc$^{-1}$. We also fixed the number of e-folds remaining until the end of inflation as $N_{\star}=55$.

We perform our analysis using the latest Planck data ~\cite{Planck:2018vyg}. Specifically, we use the high-multipole \texttt{HiLLiPoP} TTTEEE likelihood \cite{Tristram:2023haj}, add the low-multipole data from the \texttt{Commander} low-$\ell$ TT and the \texttt{LoLLiPoP} EE likelihoods. Combined with the updated CMB lensing potential likelihood \cite{Carron:2022eyg}, we denote these data sets as PR4. We also consider the 2018 BICEP/Keck Array B-mode polarization data (BK18) from the BICEP3 and Keck Array CMB Experiments \cite{BICEP:2021xfz}.

\begin{figure}[h]
\centering
\includegraphics[width=\columnwidth]{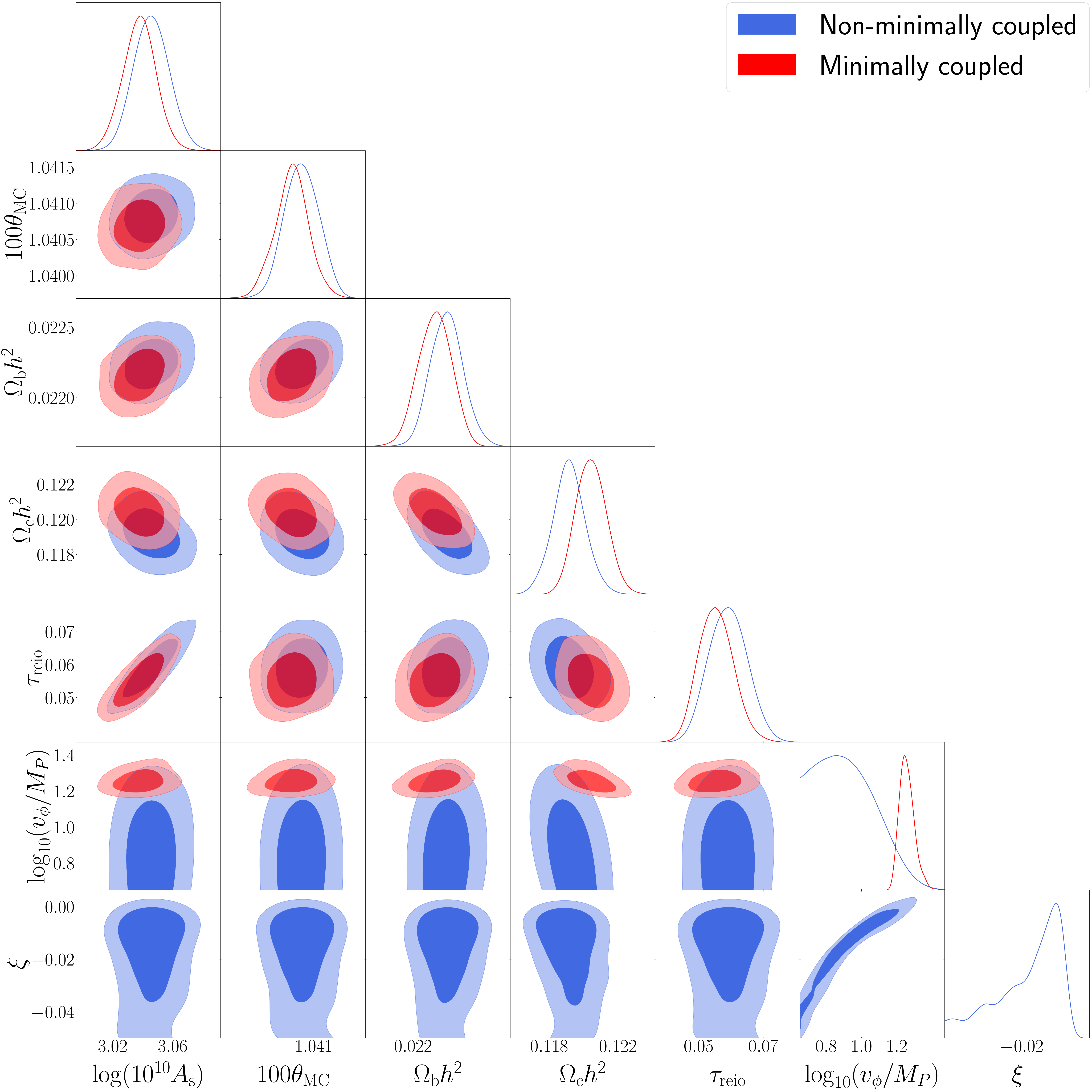}
\caption{Confidence contours at $68\%$ and $95\%$ C.L. for the minimally and non-minimally coupled double-well models, constrained using Planck+BK18 data.}
\label{fig:3}
\end{figure}

Fig. \ref{fig:3} and Table \ref{tab:1} summarize the main results of our analyses. When considering PR4+BK18 data, we obtained vacuum expectation values for the minimally and non-minimally coupled cases, respectively, of $v_\phi/M_P = 1.258^{+0.035}_{-0.048}$ and $v_\phi/M_P = 0.92^{+0.084}_{-0.26}$ at a 68\% (C.L.). We also obtain a constraint on the coupling strength for the non-minimal case, $\xi = -0.0169^{+0.015}_{-0.0059}$ also at 68\% (C.L.). This result seems to exclude completely the minimally-coupled case, in which a negative $\xi$ is preferred, in agreement with the $n_s$ and $r$ predictions shown in Fig. \ref{fig:1}. We also observe a clear correlation between the $\xi$ and $\log_{10}v_\phi/M_P$ parameters, in that a lower vev is achieved if the coupling becomes more negative.

There is a slight difference in the constraints on the inflationary parameters between the minimal and non-minimal cases, with the most significant difference concerning the vacuum expectation values. We also find a small shift towards lower values of $H_0$ in the minimally coupled scenario.

\begin{table*}[t]
		\centering
		\begin{tabular}{>{\scriptsize}c >{\scriptsize}c>{\scriptsize}c>{\scriptsize}c>{\scriptsize}c}
			\hline
			\hline
			&Minimal DW & & Non-minimal DW \\
			\hline
			  {Parameter} & Mean value & Best-fit & Mean value & Best-fit \\
            \hline 
            \multicolumn{5}{c}{\scriptsize{PR4+BK18}}\\
			
			$\omega_{b}$ &  $0.02216\pm 0.00012$ & $0.02213$ & $0.02224\pm 0.00012$ & $0.02224$ \\
			$\omega_{c}$ & $0.12044\pm 0.00089$ & $0.12031$ & $0.11905\pm 0.00096$ & $0.11984$ \\
            $100\theta_{MC}$ &  $1.04069\pm 0.00024$ & $1.04063$ & $1.04083\pm 0.00024$ & $1.0407$\\
            $\tau_{reio}$ &  $0.0554\pm 0.0054$ & $0.0519$ & $0.0591\pm 0.0059$ & $0.0619$\\
            $\ln 10^{10}A_{s}$ &  $3.038\pm 0.011$ & $3.031$ & $3.046\pm 0.012$ & $3.051$\\
			$H_0$ &  $66.96\pm 0.40$ & $66.96$ & $67.53\pm 0.42$ & $67.24$\\
			$\log_{10}v_{\phi}/M_P$  & $1.258^{+0.035}_{-0.048}$ & $1.247$ &  $0.920^{+0.084}_{-0.26}$  & $0.9287$ \\	
			$\xi$ &  $-$ & $-$ & $-0.0169^{+0.015}_{-0.0059}$ & $-0.01119$ \\
			\hline
            $\ln B$ & $-7.815$ & & $-5.689$ & \\
            \hline
            \hline 
		\end{tabular}
		\caption{The 68\% (C.L.) estimates and best-fit values for the cosmological parameters obtained from the PR4+BK18 data.}
        \label{tab:1}
	\end{table*}

For completeness, we also analyse whether the minimally or non-minimally coupled cases are preferred by the data when compared to the concordance model. We perform this analysis by using a Bayesian model-selection~\cite{Trotta:2008qt}. From the Bayes' theorem we can write the posterior probability of one model given a dataset d as,
\begin{equation}
    p(m|d)\propto p(d|m)\pi(m)\;,
\end{equation}
where $p(d|m)$ is the marginal likelihood for the model, or often called evidence $\mathcal{Z}$, and $\pi(m)$ is the prior probability of the model before the data. If we consider two models, $m_0$ and $m_1$, the \textit{Bayes factor} is given by the ratio of the evidence for the two models,
\begin{equation}
    B_{0,1}=\frac{p(d|m_0)}{p(d|m_1)}=\frac{\mathcal{Z}(m_0)}{\mathcal{Z}(m_1)}.
\end{equation}
Here, we are assuming that $\pi(m_0)=\pi(m_1)=1/2$, that is, the priors are noncommittal about the alternatives. The strength of evidence will be defined by $|\ln{B_{0,1}}|$, and is interpreted using the Jeffreys’ scale~\cite{Grandis:2016fwl}. For a $|\ln{B_{0,1}}| < 1$, we have no preference for either model, is inconclusive. For a value of $|\ln{B_{0,1}}| =1$, it shows a positive evidence for the model $m_0$ over the model $m_1$. For $|\ln{B_{0,1}}| =2.5$ and $|\ln{B_{0,1}}| = 5$, it gives a moderate and strong evidence in favors of $m_0$, respectively. 

We analyze our generated MCMC chains with the \texttt{MCEvidence} code \cite{Heavens:2017afc}. Following the latest Planck report, we compare our results with the Starobinsky model. Table \ref{tab:1} shows the Bayes factor ($\ln B$) for the considered models. We find $\ln B=-7.815$ and $\ln B=-5.689$ for the minimally and non-minimally coupled models, respectively, showing a clear preference for the Starobinsky model; however, the addition of a nonminimal coupling alleviates this result, as it helps bring the inflationary predictions to the acceptable limits, at the cost of increased complexity. There is strong evidence against these models, which is understandable since previous results on the minimally coupled model already indicated moderate evidence against it, confirmed and exacerbated by the updated data sets. Although the non-minimally coupled model alleviates this issue, the inclusion of $\xi$, which increases the parameter space, will also contribute to the computation of the evidence. Therefore, we find a result compatible with the model's added complexity.

\section{Rare Lepton Decays} \label{sec:4}

We proceed with our analysis by exploring possible consequences of the Double-Well inflation in the context of the seesaw mechanism. The canonical structure of type-I seesaw includes three sterile fermions and a scalar $SU(2)_L$ singlet to the SM content in order to introduce the Lagrangian terms,
\begin{equation}
    \mathcal{L} \supset h_{ij}\bar{f}_{iL}i \sigma_{2}H^\dagger N_{j} +h^{\prime}_{ij} \phi \bar{N}^C_{i}N_{j}, \label{YuTermI}
\end{equation}
where $f=(\nu \,\,,\,\, e)_L^T$ and $H=(H^+ \,\,,\,\, H^0)^T$ are the standard leptons and the Higgs doublets, respectively. By allowing a non-zero vev for the singlet field one obtains a Majorana mass term for the sterile fermions, $M = h^{\prime}\, v_\phi/\sqrt{2}$. Here, $v_\phi$ is constrained by cosmological perturbation data (Table \ref{tab:1}). A Dirac mass term is also obtained once the electroweak symmetry breaking is accomplished, $m_D = h\, v_h/\sqrt{2}$. Once the block diagonalization procedure \cite{Schechter:1981cv,Hettmansperger:2011bt,Dias:2012xp} has been applied, it is possible to use the information on the neutrino oscillation pattern (mixing angles and quadratic mass differences~\cite{Workman:2022ynf}) to estimate their Yukawa couplings, $h_{ij}$ and $h^\prime_{ij}$. The latter is particularly useful when estimating extra contributions to flavor violating rare lepton decays.
 
Such processes can be triggered through active-sterile neutrinos superposition in the charged current Lagrangian,
\begin{eqnarray}
    \mathcal{L}_{CC} &&= -\frac{g}{\sqrt{2}}\bar{e}_{iL} \gamma^\mu \nu_{iL} W^-_\mu + h.c. \nonumber \\
    && \simeq -\frac{g}{\sqrt{2}}\bar{e}_{iL} \gamma^\mu \left\{U^0_{ij}n^0_{jL} + F_{ij}n^1_{jL} \right\}W^-_\mu + h.c.\, ,
\end{eqnarray}
where $F = m_D M^{-1}$ and $U^0$ is the Pontecorvo-Maki-Nakagawa-Sakata (PMNS) matrix. The contribution of the sterile (heavy) neutrinos to the charged current is suppressed by the ratio $v_h/v_\phi$. We consider the predictions of the model only to the decay channel $\mu \rightarrow e \, \gamma $, as the current observational bound to this process is the most restrictive one~\cite{Workman:2022ynf}. 

The branching ratio for the decay mediated by the three sterile neutrinos is evaluated by~\cite{Ilakovac:1994kj}
\begin{equation}
    B(\mu \rightarrow e \, \gamma) = \frac{\alpha^3_W \sin^2(\theta_W)}{256\pi^2}\frac{m^5_\mu}{m^4_W \Gamma_{\mu}}\left|\sum^3_{i=1} F_{2i} F_{1i}\,I\left( \frac{M^2}{m^2_W}\right) \right|^2,
\end{equation}
where
\begin{equation}
    I(x) = -\frac{2x^3 + 5x^2 - x}{4(1-x)^3} - \frac{3x^3\ln{x}}{2(1-x)^4}\;.
\end{equation}
In the expression above, $\alpha_W$ stands for $\frac{g^2}{4\pi}$ and $g$ is the weak gauge coupling. $\theta_W$ is the Weinberg angle, while $\Gamma_{\mu}$ is the total decay width of the muon. $m_\mu$ and $m_W$ are the pole mass for the muon and the $W$ boson, respectively. The values for these quantities can be found in~\cite{Workman:2022ynf}.  We also consider the normal hierarchy case and a degenerated sterile sector with   $h^\prime \sim 10^{-4}$\footnote{$h^\prime \lesssim 10^{-4}$ in order to preserve the unitary aspect of neutrino-neutrino scattering.}, the best-fit value from the MCMC analysis for the inflaton vacuum $v_\phi \simeq 17.66\,M_P$ and the electroweak breaking scale  $v_H \simeq 246$ GeV to obtain 
\begin{equation}
    B(\mu \rightarrow e \, \gamma) \simeq 3.93 \times 10^{-53},
\end{equation}
which is negligible when compared to the highly restrictive superior bound on $B(\mu \rightarrow e \, \gamma) < 4.2 \times 10^{-13}$~\cite{Workman:2022ynf}. 

The same process can be employed to obtain the corresponding branching ratio in the non-minimal scenario of the Double-Well inflation. In this case, one needs to re-scale the Dirac and Majorana mass terms according to the Weyl factor $\Omega^2 = 1+\xi\, \phi^2/M^2_P$, $\tilde{m}^D =m^D/\Omega$ and $\tilde{M} =M/\Omega$, following the conformal transformation of the metric to the Einstein frame \cite{Garcia-Bellido:2008ycs}. For $h^\prime = 10^{-4}$ and considering the best-fit value for $\xi \simeq -0.011$ and $v_\phi = 8.49\, M_P$\footnote{One should be aware that some combinations of the parameters $\xi$ and $v_\phi$ can lead to a negative conformal factor, $\Omega^2<0$, even within the $1\sigma$ confidence limit for their integrated posterior probabilities. The distorted 2D confidence region obtained for these parameters is a symptom of these nonphysical combinations.}, one obtains $ B(\mu \rightarrow e \, \gamma) \simeq 3.31 \times 10^{-53}$.

\begin{figure}[t]
\centering
\includegraphics[width=\columnwidth]{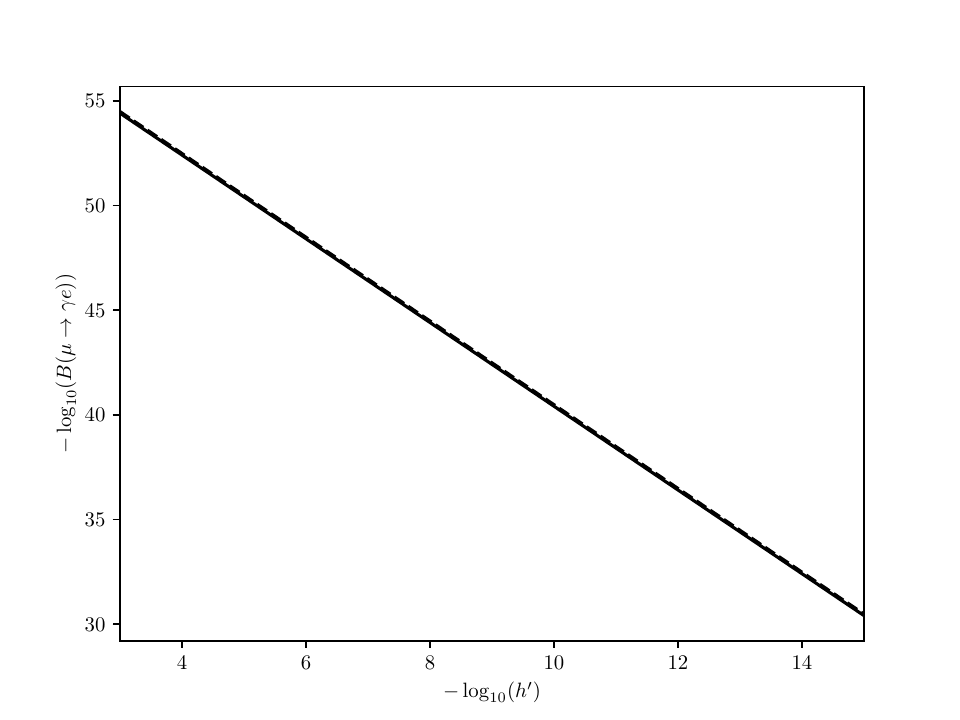}
\caption{Branching ratio of the rare lepton decay $\mu \rightarrow e \, \gamma$ against the Yukawa coupling $h^\prime$ for the minimal (solid black) and non-minimal (dashed black) scenarios of the Double-Well inflation. We consider $\sum m_i = 0.057$ eV.}
\label{fig:Rare}
\end{figure}

Fig.~\eqref{fig:Rare} shows the branching ratio in the minimal and non-minimal scenarios for different values of $h^\prime$, where it is evident that a sizable contribution to the Lepton violating decay is obtained only for tiny values of the Yukawa coupling of the sterile neutrinos, roughly $h^\prime \lesssim 10^{-20}$. Considering that one of the goals of the seesaw mechanism is to obtain the correct neutrino mass pattern without resourcing to huge fine-tuning in the Yukawa sector, one may infer that such a signature of the seesaw mechanism is highly improbable to be detected, relegating the sterile neutrinos of the Double-Well scenario to the dark sector of the Universe. The decay in the nonminimal scenario is slightly smaller when compared to the minimal one.

\section{Discussion \label{conclusion}} \label{sec:5}

The double-well potential arises in the context of spontaneous symmetry breaking, where the vacuum expectation value $v_\phi$ determines the energy scale of the transition from a symmetric to an asymmetric phase of the system. Once settled in the vacuum state, the scalar field acquires a positive non-zero squared mass term $m^2_\phi = \lambda v^2_\phi$.  This transition is not instantaneous in general, allowing for a slow-roll phase at the cost of a rather unnatural large vev.

In this paper, we performed a complete statistical analysis of the observational viability of the small-field Double-Well inflationary model. In particular, we applied a Bayesian procedure of parameter selection and comparison of models to probe the likelihood of the model against the Starobinsky scenario, as well as to investigate the phenomenological consequences of the underlying field theory. From the comparison of models, we obtained a preference for the reference (Starobinsky) model against the minimally coupled double-well model, in agreement with results discussed in the Planck 2015 report~\cite{planck15}, but now with a more significant difference due to the updated data sets. Including a non-minimal coupling with gravity improves the results, but the higher complexity still makes the model less favorable.

For the scenario minimally coupled to gravity, we obtain $v_\phi \simeq 17.66\,M_P$ according to the selection of parameters procedure (best-fit value). Once the Planck normalization to the amplitude of scalar perturbation is considered, one can compute the inflaton’s quartic coupling $\lambda \simeq 2.86 \times 10^{-15}$,  which enables $m_\phi \simeq 9.45 \times 10^{-7}\,M_P$ for the mass of the inflaton field. 

In the non-minimal scenario these parameters are slightly different, following the effect of the direct coupling between the inflaton field and the Ricci scalar.  Assuming $v_\phi \simeq 8.49\, M_P$ for the vev, $\lambda \simeq 3.28 \times 10^{-13}$ for the inflaton's coupling and $\xi \simeq -0.011$ for the non-minimal coupling strength, one should obtain for the mass of the inflaton $m_\phi \simeq 1.10 \times 10^{-5}\,M_P$.

Such high values for the inflaton mass prevent this particle from being produced thermally in the early universe, as well as produced on-shell by collider experiments. One may be tempted to investigate the secondary effects triggered by the inflaton field immersed in a context beyond the cosmological set up. We investigated the consequences for the inflaton field immersed in the framework of canonical type-I seesaw and computed the contributions of the right-handed neutrinos for the rare lepton decay $\mu^- \rightarrow e^-\, \gamma$. We found such a contribution to be negligible when compared to the current experimental upper bound on this processes. 

The large values inferred for the inflaton's vev in the Double-Well scenarios seclude the associated sector to complete darkness. Rescuing the Double-Well models for the visible sector must necessarily cover a realization of the inflation period for lower values of $v_\phi$. A particularly attractive scenario may arise in the framework of warm inflation \cite{Berera:1995ie,Berera:2008ar}, where the dissipation coefficient over-dampens the inflaton's trajectory in field space, optimizing the slow-roll. The predicted results of this configuration will appear in a forthcoming communication.


\section*{Acknowledgements}
GR is supported by the Coordena\c{c}\~ao de Aperfei\c{c}oamento de Pessoal de N\'ivel Superior (CAPES). JGR and FBMS acknowledge financial support from the Programa de Capacita\c{c}\~ao Institucional do Observat\'orio Nacional (PCI/ON/MCTI). JSA is supported by CNPq grant No. 307683/2022-2 and Funda\c{c}\~ao de Amparo \`a Pesquisa do Estado do Rio de Janeiro (FAPERJ) grant No. 259610 (2021). We also acknowledge the use of \texttt{cobaya}, CAMB and \texttt{MCEvidence} codes. This work was developed thanks to the use of the National Observatory Data Center (CPDON).

\bibliography{bibliography}

\label{lastpage}

\end{document}